\documentstyle[11pt,newpasp,twoside,epsf]{article}
\def\edcomment#1{\iffalse\marginpar{\raggedright\sl#1\/}\else\relax\fi}
\reversemarginpar

\def\p{\partial}

\def\nab{\mbox{\boldmath $\nabla$}}
\def\rb{\bar{\rho}}
\def\tb{\bar{T}}
\def\sb{\bar{S}}

\def\vph{\hat{v}_{\phi}}

\begin{document}

\title{Solar Turbulence and Magnetism Studied Within a Rotating Convective Spherical Shell}
 \author{Allan Sacha Brun \& Juri Toomre}
\affil{JILA, University of Colorado, Boulder, CO 80309-0440, USA.}
\begin{abstract}
We discuss recent advances made in modelling the complex magnetohydrodynamics 
of the Sun using our anelastic spherical harmonics (ASH) code. 
We have conducted extensive 3--D simulations of compressible convection
in rotating spherical shells with and without magnetic fields, to study the 
coupling between global-scale convection and rotation in seeking to understand 
how the solar differential rotation is established and maintained.  
Such simulations capable of studying fairly turbulent convection have been enabled 
by massively parallel supercomputers. The resulting
convection within domains that capture a good fraction of the bulk of
the convection zone is highly time dependent and intricate, and is
dominated by intermittent upflows and networks of strong downflows.  A
high degree of coherent structures involving downflowing plumes
can be embedded in otherwise chaotic flow fields. These vortical structures
play a significant role in yielding Reynolds stresses that serve to
redistribute angular momentum, leading to differential rotation
profiles with pole to equator contrasts of about 30\% in angular velocity 
$\Omega$ and some constancy along radial line at mid latitudes, thereby making 
good contact with deductions from helioseismology. When a magnetic field 
is introduced, a dynamo regime can be found that does not destroy the strong 
differential rotation achieved in pure hydrodynamics cases. 
The magnetic fields are found to concentrate around the downflowing networks
and to have significant north-south asymmetry and helicity. 
\end{abstract}

\section{Observational Challenges}

The surface layers of the Sun have long been known to exhibit complex convection 
and magnetism involving a very broad range of spatial and temporal scales (Stix 2002).
Helioseismology is now permitting novel views of the interior structure and dynamics
within our nearest star (Gough \& Toomre 1991). Using millions of acoustic modes, 
it is possible to probe as a function of radius and latitude the solar sound 
speed $c$, density $\rho$ and angular velocity $\Omega$. Figure 1 shows the solar internal 
rotation profile as a contour plot and as six latitudinal cuts (Schou et al. 1998, 
Howe et al. 2000). Although there are prominent variations of $\Omega$ with latitude, 
with the equator rotating considerably faster than the high latitudes, it is quite 
striking that $\Omega$ is largely constant at mid latitude along radial lines 
and imprints itself all the way down to the base of the convective zone. 
There a tachocline of strong velocity shear joins the nearly solid body rotation 
of the radiative interior with the differentially rotating convective zone. A distinctive 
near-surface shear layer is also evident. Such angular velocity patterns were not anticipated 
and are at variance with earlier models of differential rotation within convecting rotating shells 
(Glatzmaier \& Gilman 1982, Gilman \& Miller 1986, Glatzmaier 1987). In those models 
$\Omega$ was mostly constant along cylinders aligned with the rotation axis (e.g. Taylor columns).

\begin{figure}[!ht]
\setlength{\unitlength}{1.0cm}
\begin{picture}(5,5.5)
\includegraphics{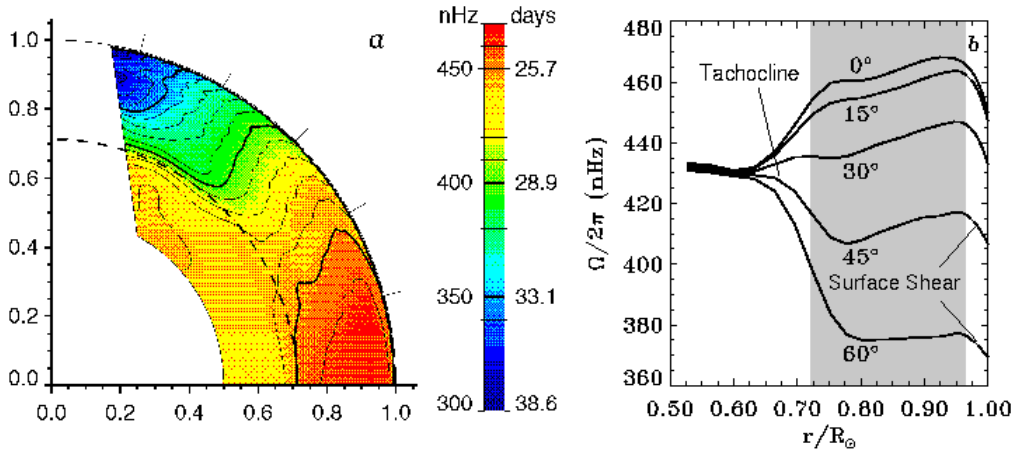}
\end{picture}
\caption{(a) Angular velocity profile $\Omega/2\pi$ in latitude and radius as deduced 
through inversion of acoustic mode frequency splittings from the SOI-MDI helioseismic 
instrument; the equator here coincides with the horizontal axis [adapted from Schou et al. 1998]. 
(b) Time-averaged rotation rates from five years of GONG helioseismic data, 
plotted against proportional radius at different latitudes, with rapid rotation at 
the equator and slower rotation at high latitudes. The zone covered by our computational 
domain is indicated (grey area) [adapted from Howe et al. 2000].}
\end{figure}

Intimately related to the dynamics of the solar turbulent convection zone is its 
magnetic activity, variously involving sunspots, prominences and CME's, along with its 
overall 22-year cycle. How such a turbulent and complex system as the Sun can exhibit 
order amidst what is seemingly chaos is a most challenging question. It is generally 
thought that the solar magnetic dynamo operates at two differing ranges of spatial and 
temporal scales (Cattaneo \& Hughes 2001). The global dynamo yielding the regular 22-year cycle and 
butterfly diagrams for sunspot emergence is likely to be seated within the tachocline 
at the base of the convection zone (Parker 1993). The origin of the rapidly varying 
and smaller scale magnetism is probably due to local dynamo action achieved by 
the intensely turbulent convection. We would like here to cast new light on some aspects of 
this complex magnetohydrodynamical (MHD) system. We believe that 3--D MHD numerical simulations 
are essential to pursue questions of solar magnetism. Given the large range of temporal and 
spatial scales involved in the solar convective envelope, one has to choose between either 
a local high resolution domain or a global and somewhat less resolved spherical domain. 
The advantage of the former is its ability to resolve highly turbulent flows 
(Brummell et al. 1998, 2002; Stein \& Nordlund 1998). The global approach, by fixing 
the largest scales, is more limited in the turbulence levels that can be resolved, 
yet takes into account the correct geometry and its topological implications for mean flows.
We will focus our attention here on the establishment of the solar differential rotation 
and meridional circulations, both of which are the purview of global models. We shall present 
recent results from our simulations of solar convection within full 3--D spherical shells, 
discussing also the angular velocity $\Omega$ profiles that can be achieved in the bulk of 
the convection zone and the level of dynamo induced magnetism that can be sustained there.

\section{Our Numerical Approach}

The ASH code solves the 3--D MHD anelastic equations of motion in a rotating spherical shell 
geometry (Clune et al. 1999, Miesch et al. 2000). These equations are fully nonlinear in velocity 
and magnetic field variables; the thermodynamic variables are separated with respect to a spherically 
symmetric and evolving mean state having a density $\rb$, pressure $\bar{P}$, temperature $\tb$ 
and specific entropy $\sb$, and fluctuations about this mean state, namely $\rho$, $P$, $T$, $S$: 

\begin{eqnarray}
\nab\cdot(\rb {\bf v}) &=& 0, \\
\rb \left(\frac{\p {\bf v}}{\p t}+({\bf v}\cdot\nab){\bf v}+2{\bf \Omega_o}\times{\bf v}\right) 
 &=& -\nab P + \rho {\bf g} + \frac{1}{4\pi} (\nab\times{\bf B})\times{\bf B} \nonumber \\
&-& \nab\cdot\mbox{\boldmath $\cal D$}-[\nab\bar{P}-\rb{\bf g}], \\
\rb \tb \frac{\p S}{\p t}=\nab\cdot[\kappa_r \rb c_p \nab (\tb+T)&+&\kappa \rb \tb \nab (\sb+S)]\nonumber \\
-\rb \tb{\bf v}\cdot\nab (\sb+S)+\frac{\eta}{4\pi}(\nab\times{\bf B})^2&+&2\rb\nu\left[e_{ij}e_{ij}-
1/3(\nab\cdot{\bf v})^2\right],\\
\frac{\p {\bf B}}{\p t}=\nab\times({\bf v}\times{\bf B})&-&\nab\times(\eta\nab\times{\bf B}),
\end{eqnarray}
where ${\bf v}=(v_r,v_{\theta},v_{\phi})$ is the local velocity in spherical coordinates in 
the frame rotating at constant angular velocity ${\bf \Omega_o}$, ${\bf g}$ is the 
gravitational acceleration, {\bf B} is the magnetic field, $c_p$ is the specific heat at 
constant pressure, $\kappa_r$ is the radiative diffusivity, $\eta$ is the effective magnetic 
diffusivity, and ${\bf \cal D}$ is the viscous stress tensor, involving the components
\begin{eqnarray}
{\cal D}_{ij}=-2\rb\nu[e_{ij}-1/3(\nab\cdot{\bf v})\delta_{ij}],
\end{eqnarray}
where $e_{ij}$ is the strain rate tensor, and $\nu$ and $\kappa$ are effective eddy diffusivities. 
To complete the set of equations, we use the linearized equation of state
\begin{equation}
\frac{\rho}{\rb}=\frac{P}{\bar{P}}-\frac{T}{\tb}=\frac{P}{\gamma\bar{P}}-\frac{S}{c_p},
\end{equation}
where $\gamma$ is the adiabatic exponent, and assume the ideal gas law 
\begin{eqnarray}
\bar{P}={\cal R} \rb \tb
\end{eqnarray}
where ${\cal R}$ is the gas constant.

The velocity and magnetic fields and the thermodynamic variables are expanded in spherical 
harmonics for their horizontal structure and in Chebyshev polynomials for their radial structure. 
This approach has the advantage that the spatial resolution is uniform everywhere on a sphere 
when a complete set of spherical harmonics is used up to some maximum in degree $\ell$ 
(retaining all azimuthal orders $m$). The anelastic approximation captures the effects of density 
stratification without having to resolve sound waves which would severely limit the time steps. 
We use a toroidal and poloidal decomposition that enforces the mass flux and the magnetic field 
to remain divergence free.

The model is a highly simplified description of the solar convection zone: solar values are 
taken for the heat flux, rotation rate, mass and radius, and a perfect gas is assumed since 
the upper boundary of the shell lies below the H and He ionization zone. The computational 
domain extends from 0.72 $R_\odot$ to 0.96 or 0.98 $R_\odot$, thereby concentrating on the 
bulk of the unstable zone and here not dealing with penetration into the radiative interior. 
The effects of the steep entropy gradient close to the surface has been softened by introducing 
a subgrid scale (SGS) transport of heat to account for the unresolved motions, and enhanced 
eddy diffusivities are used in these large eddy simulations (LES). The typical density difference 
across the shell in radius is about 30.

In order to get reliable statistics, these numerical experiments need to be integrated over
long periods in physical time. Within our 3--D convection simulations the dynamical time is of 
the order of 30 days (the time roughly taken by a fluid element to travel across the convection zone). 
This implies that time averages of relevant quantities such as $\Omega$ have to be performed over 
no less than 300 days when feasible. We here present one of our most turbulent cases, namely case $E$ 
computed with a Prandtl number $P_r=\nu/\kappa$ of 0.25 and a rms Reynolds number $R_e=v_{rms}D/\nu$ 
of $\sim$ 650, where $D$ is the thickness of the shell. This case has been restarted from a 
promising laminar case $AB$ (Brun \& Toomre 2002) which had been evolved over 200,000 time steps  
for a total physical time of about 6200 days starting from quiescent initial conditions. To 
achieve more turbulent states such as in case $E$, we have lowered the viscous and thermal 
effective diffusivites $\nu$ and $\kappa$ in case $AB$ gradually in  a sequence of steps, 
waiting for the simulation to relax between each step. For case $E$ we present the latest 85 days of the 
simulation over a total of 9,200 physical days of evolution (corresponding to times steps 430,000 
to 480,000). At this level of turbulence, numerical accuracy requires a resolution of 
$\ell_{max}=680$ and a physical time step of 150 s. These numbers illustrate the difficulty 
of conducting such 3--D turbulent simulations. Although our ASH code 
is sustaining 300 Mflops/cpus or 20\% of peak performance, thus performing very well on massively 
parallel supercomputers such as the IBM-SP3 at SDSC, such  a single simulation requires of order
80,000 cpu hours. Typical runs use 256 processors, which is equivalent to 76 Gflop/s. We now describe in
details some results from our 3--D convection simulations.

\section{Turbulent Convection under the Influence of Rotation}

Figure 2 displays the evolution of radial velocity in case $E$ over 10 days in time near 
the top of the domain. Dark tones represent downflows and lighter ones upflows. The vantage 
point is in the uniformly rotating frame used in our simulations. The convection patterns 
are intricate and highly time dependent. Some of the pattern evolution is related to 
the advection by prograde zonal flows near the equator and retrograde ones at higher latitudes 
associated with the differential rotation driven by the convection relative to this frame. 
Shearing and cleaving of the convective cells is evident, as well as distortions in the downflow lanes. 
There is an asymmetry between the broad upflows centered in each convecting cell and 
the narrow fast downflows at their periphery. This leads to a downward
transport of kinetic energy. The strong correlations between warm upward motions and 
cool downward motions are essential in transporting the heat outward. 

\begin{figure}[!ht]
\setlength{\unitlength}{1.0cm}
\begin{picture}(5,4.2)
\includegraphics{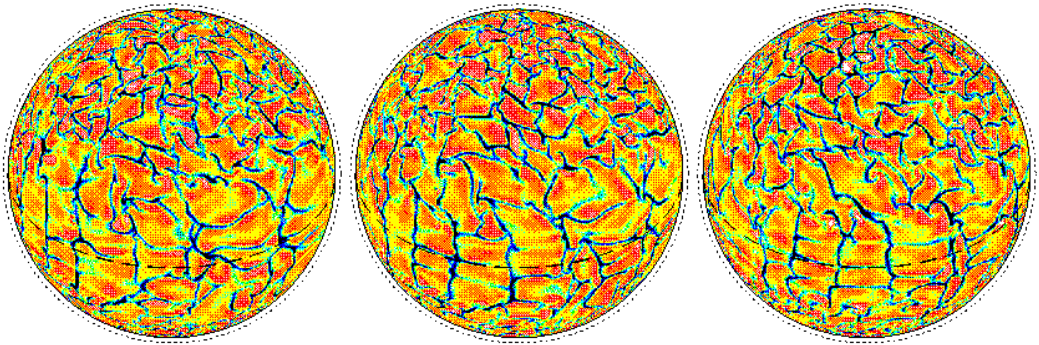}
\end{picture}
\caption[]{\label{fig2} Evolution in the convection over 10 days, showing the 
radial velocity in case $E$ near the top (0.97 $R_\odot$) of the spherical domain. 
The time interval between each successive image is about 5 days. Downflows appear dark.
The dotted circle is located at the solar radius $R_\odot$ and the equator is indicated
by the dashed curve.}
\end{figure}

Pronounced vortical structures are evident at the interstices of the downflows network.
They are counterclockwise in the northern hemisphere and clockwise in the southern one, 
i.e cyclonic. The strongest of these vortex tubes or `plumes' extend through the whole domain depth. 
These plumes represent coherent structures that are surrounded by more chaotic flows. 
They tend to align with the rotation axis and to be tilted away from the meridional planes,  
leading to Reynolds stresses that are crucial ingredients in redistributing the 
angular momentum within the shell.

\subsection{Making Good Contact with the Solar Differential Rotation}

\begin{figure}[!ht]
\setlength{\unitlength}{1.0cm}
\begin{picture}(5,6.3)
\includegraphics{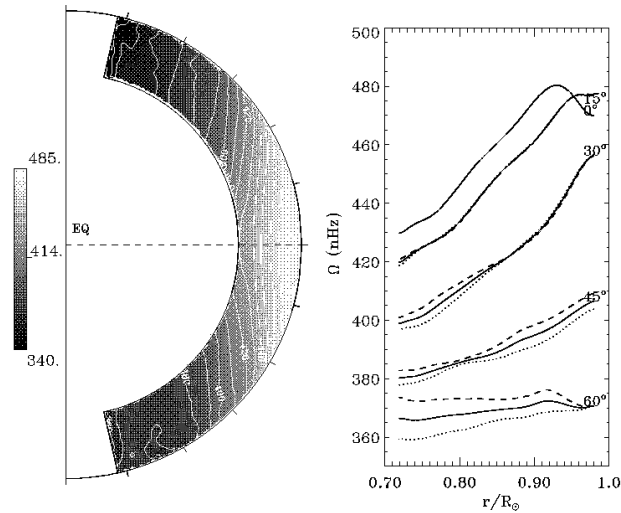}
\end{picture}
\caption[]{\label{fig3} Temporal and longitudinal averages in cases $E$ of the angular velocity 
profiles formed over an interval of 85 days. This case exhibits a prograde
equatorial rotation and a strong contrast $\Delta \Omega$ from equator to pole, as well as
possessing a high latitude region of particularly slow rotation. In the right panel a sense
of the asymmetry present in the solution can be assessed in these radial cuts at indicated latitudes.}
\end{figure}

The differential rotation profile in latitude and radius associated with the vigorous 
convection of case $E$ is shown in Figure 3. For simplicity, we have converted the mean 
longitudinal velocity $\vph$ into a sidereal angular velocity $\Omega$, using 
$\Omega_o/2\pi=414$ nHz (or 28 days) as the reference frame rotation rate. 
In the contour plot, the near polar regions have been omitted due to the difficulty 
of forming stable averages there, since the averaging domain is small
but the temporal variations large. Case $E$ exhibits a fast (prograde) equatorial
region and slow (retrograde) high latitude region. This is due to correlations in the velocity 
components leading to significant Reynolds stresses. These Reynolds stresses are intimately 
linked to the influence of Coriolis forces acting upon the convecting motions and 
to the presence of plumes tilted both away from the local radial direction and out 
of the meridional plane. Such correlations have been identified in local high resolution
Cartesian domains as well (Brummell et al. 1998). These lead to an equatorward transport 
of angular momentum, resulting in the slowing down of the high latitude regions and speeding  
up of the equatorial zone. At low latitudes there is some alignment of $\Omega$ along the rotation axis. 
At mid latitudes, the angular velocity is nearly constant along radial lines, in good agreement 
with helioseismic deductions (cf. Fig. 1). Further, case $E$ exhibits a monotonic decrease 
of $\Omega$ with latitude, a property that has been difficult to achieve in 3--D spherical 
convection calculations. Indeed, most other cases have their equator to pole contrast 
$\Delta\Omega$ confined to mid latitudes (i.e where the inner tangent cylinder cuts through the 
outer shell at $\sim 42^\circ$). The differential rotation contrast between the equator and 
$60^\circ$ in case $E$ is 110 nHz (or 26\% relative to the frame of reference), 
thus being very close to the 92 nHz (or 22\%) variation observed in the Sun. Since our 
progenitor case $AB$ shared this attribute as well, it is comforting that we can retain 
this solar-like property in a significantly more turbulent and complex case such as case $E$. 
Many of the more complex cases that we have computed previously had a 
tendency to lose some of their latitudinal contrast in $\Omega$,  which also became 
more nearly constant along cylinders aligned with the rotation axis (Brun \& Toomre 2002). 
A sense of the asymmetry present in case $E$ can be assessed both in the contour plot 
and in the latitudinal cuts (right panel of Fig. 3), where we have plotted $\Omega$ 
in the north (dotted) and south (dashed) hemispheres along with their mean. 
The convection itself exhibits some asymmetry between the two hemispheres (cf. Fig. 2), and 
so it is not surprising that the mean flows driven by the convection do the same. These asymmetries
are expected to diminish over a longer temporal average. 
Mean field models of the solar differential rotation (Kichatinov \& R\"udiger 1995, Durney 1999) 
have advocated that a thermal wind balance (involving pole to equator temperature
contrasts) could be the cause of the non-cylindrical profile in $\Omega$.
This could come about through the baroclinic nature of the convecting motions yielding some
latitudinal heat fluxes, resulting in the breakdown of the Taylor-Proudman theorem (Pedlosky 1987). 
Although it is indeed true that case $E$ exhibits latitudinal 
variation of entropy and temperature fluctuations relative to the mean, these are not
the most dominant players everywhere in the shell. A temperature contrast of few degree K seems
compatible with a $\Delta\Omega/\Omega_o$ of $\sim 30\%$. However, we find that the 
Reynolds stresses are the main agents responsible for the equatorial acceleration 
achieved in our simulations, and thus the solar differential rotation is dynamical in origin. 

\subsection{Pending Issues Concerning the Meridional Circulation}

The meridional circulation associated with the vigorous convection in case $E$ is 
maintained variously by Coriolis forces acting on the differential rotation, by buoyancy forces, 
by Reynolds stresses and by pressure gradients, and thus can be thought as a small departure from geostrophic 
balance. The meridional circulation exhibits a multi-cell structure both in latitude and radius, 
and given the competing processes for its origin, it is not straightforward to predict. 
Typical amplitudes for the velocity are of order 25 {\rm m/s}, comparable to local helioseismic 
deductions (Haber et al. 2002). The flow is directed poleward at low latitudes, with return flow 
deeper down. The temporal fluctuations in the meridional circulation are large and thus stable  
time averages are only attained by sampling many rotations. The kinetic energy in the 
differential rotation and in the  convective motions are two orders of magnitude higher 
than that in the meridional circulation (Brun \& Toomre 2002). As a result, small 
fluctuations in the convective motions and differential rotation can lead to major 
variations in the circulation. Some of the helioseismic inferences suggest 
the presence of single cell circulations, which are at odds with our multi-cell patterns. 
However these inferences vary from year to year, and there is recent evidence for double-cell 
structure in the circulations observable in the near-surface shear layer, but only 
in the northern hemisphere as the current solar cycle advances (Haber et al. 2002). 
From a careful analysis of the angular momentum transport in our shell we have deduced 
that the slow pole behavior seen in case $E$ seems to come about from a relatively 
weak meridional circulation at high latitudes. Case $E$ shares with case $AB$ the 
property of relatively mild meridional circulation at the higher latitudes. 
This permits a more efficient extraction of angular momentum by the Reynolds stresses 
from the high latitudes toward the equator in yielding the interesting differential
rotation profile that is achieved. 

\section{Adding Magnetic Fields to the Solar Cauldron}

We now turn to consider the influence of magnetic fields both upon the convection in our 
deep shell and upon the angular velocity profiles that can be maintained.
Early attemps to explain the 22-year solar cycle considered the possibility that the solar 
dynamo operated within the bulk of the convective envelope (Gilman 1983, Glaztmaier 1985), 
but such approaches failed because strong magnetic fields could not be stored efficiently
within the unstable stratification of the convection zone. The dynamo periods were too short 
(of the order of 1 year) and the poloidal fields were found to propagate poleward, at variance with 
observations. More recently, Parker (1993) has proposed an interfacial dynamo model
seated in the stable tachocline (see also R\"udiger \& Brandenburg 1995). 
Magnetic field is still generated within the bulk of the convection zone, but is pumped downward
into the stable layer via overshooting turbulent plumes, to be stretched there into 
large-scale toroidal structures by the strong shear of the tachocline. When the 
amplification of the toroidal magnetic field is great enough, the structures (called
magnetic flux tubes in mean field models) become magnetically buoyant and rise upward through
the convective envelope. The strongest of those structures emerge in
the photosphere as bipolar magnetic arcades, whereas the weaker ones are recycled 
within the convective zone. This leads to the crucial natural cycle of poloidal 
to toroidal interchange, i.e. $B_{pol} \rightarrow B_{tor} 
\rightarrow B_{pol}$. However, many aspects of these essential `dynamo building blocks' 
remain to be demonstrated through nonlinear 3--D calculations. At present it is not 
feasible to simulate self consistently all the processes operating together, and thus 
one needs to concentrate on individual components (magnetic generation, pumping, 
shearing, buoyant rising). One important ingredient in the interfacial dynamo scenario is the ability
of the convective motions to generate and sustain magnetic fields in the bulk of the zone. 
We thus have evaluated some conditions in 3--D compressible convection for which such a dynamo 
threshold can be realized. We further wished to identify the maximum nonlinear amplification of the
magnetic field that can be sustained by the convective motions without 
destroying the strong angular velocity contrasts previously attained.

\begin{figure}[!ht]
\setlength{\unitlength}{1.0cm}
\begin{picture}(5,7)
\includegraphics{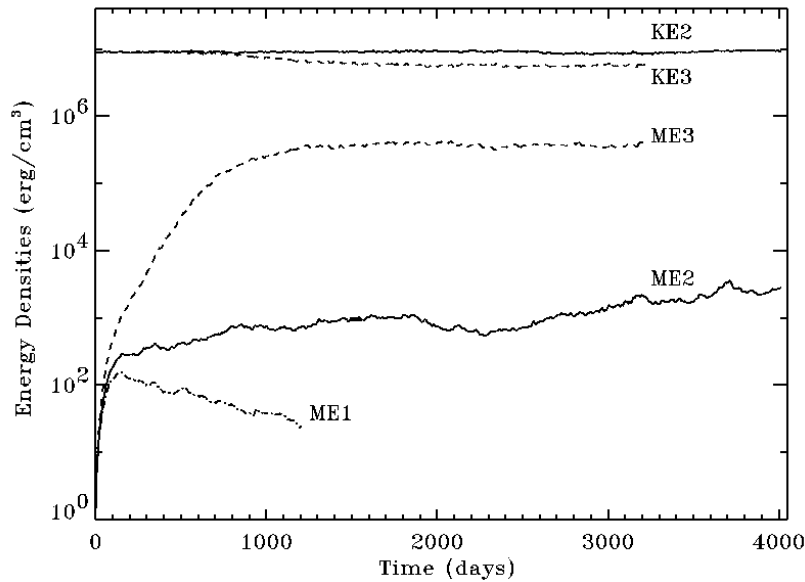}
\end{picture}
\caption[]{\label{fig4} Kinetic energy (KE) and magnetic energy (ME) for cases $M1$, $M2$ and $M3$,
involving in turn a magnetic rms Reynolds number $R_{em}$ of 250 (dashed dot line), 300 
(solid) and 500 (dashed).}
\end{figure}

\subsection{Nonlinear Dynamo Threshold and MHD Turbulence}

We have conducted three MHD simulations (named respectively $M1$, $M2$ and $M3$) started from a
solution slightly more turbulent than case $AB$ but exhibiting a similar angular velocity 
profile $\Omega$. We then introduced a small seed dipolar magnetic field and let the simulations 
proceed. Because we had to solve two extra equations (i.e the poloidal and toroidal components
of the induction equation [4]) and integrate these solutions over several ohmic
diffusion time, it was not feasible to use our more turbulent case $E$ that required higher
spatial resolution. Figure 4 shows the magnetic energy evolution for three values of the 
magnetic diffusivity $\eta$ (i.e. $2\times 10^{12}$, $1.6\times 10^{12}$, $10^{12}$ 
{\rm cm$^2$/s}). We note that over more than 3000 days (corresponding to several 
ohmic decay times) the two lowest diffusive cases $M2$ and $M3$ exhibit a sustained 
magnetic energy (ME), the levels of which depend on $\eta$. 
The other case $M1$ is clearly decaying, since the rate of generation of magnetic fields  
could not compensate for the rate of destruction by ohmic diffusion. 
The dynamo threshold seems to be around a magnetic Reynolds number $R_{\rm em}=v_{\rm rms}D/\eta$ of 
$\sim 300$ or $\eta\sim 1.6\times 10^{12}$. This is about 25\% higher than in a progenitor  
incompressible simulation (Gilman 1983).

\begin{figure}[!ht]
\setlength{\unitlength}{1.0cm}
\begin{picture}(5,6.2)
\includegraphics{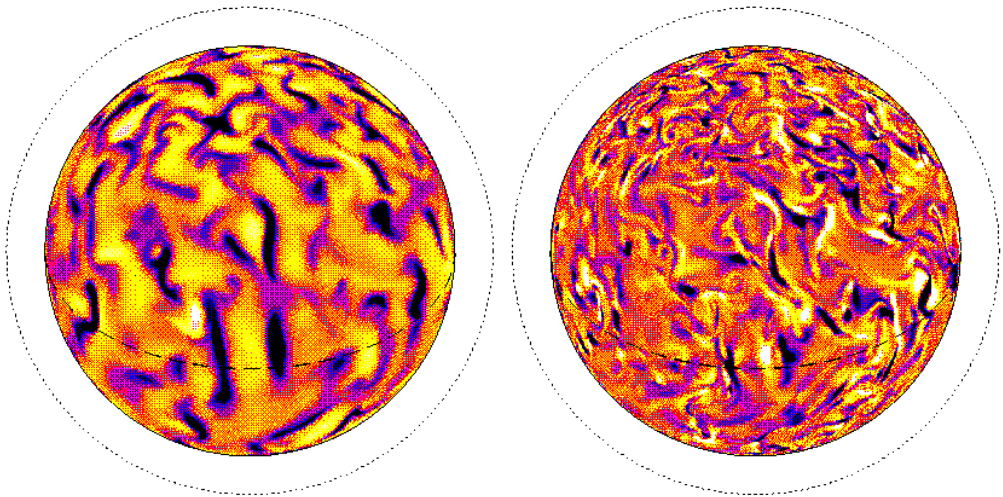}
\end{picture}
\caption[]{\label{fig5} Snapshots of the radial component of the velocity (left) and
magnetic (right) fields in case $M3$ in the middle of the domain (0.86 $R_\odot$). Dark tones
represent downflow (or negative polarity). The velocity and magnetic field peak amplitudes are 
about 100 {\rm m/s} and a few thousand Gauss respectively. The outer dotted circle 
corresponds to solar radius $R_\odot$ and the dashed curve indicates the equator.}
\end{figure}

Further, we note that the kinetic energy (KE) in model $M3$ has been reduced by about 40\% compared to
its initial value. In this case ME has grown to reach a
value of 7\% of KE. The increase of the magnetic energy started to influence the total amount of 
kinetic energy contained in the shell when ME reached roughly 0.5\% of KE after about 600 days of
evolution. The early exponential growth of ME in case $M3$ extends to about 600 days from
the start, after which the nonlinear feedback of the Lorentz forces on the flow begins to be felt. 
For case $M2$, ME is still small enough  (i.e $\leq 0.1\%$) even after 
4000 days for the convective motions to only be mildly affected by the Lorentz forces. 
This is most clearly seen in comparing the kinetic energy time trace for cases $M2$ and $M3$.
Figure 5 displays snapshots of the radial component of both the velocity and the magnetic 
fields for case $M3$ in the middle of the domain. The magnetic field is mainly 
concentrated in the downflow lanes, having been swept away from the center of the convective cells 
by the broad upflows. Both polarities coexist at the downflow network interstices. 
Clearly the magnetic field has a finer and more intricate structures, exhibiting
many swirls. This is due to our choice of magnetic diffusivity being the smaller,
noting that the magnetic Prandtl number $P_{rm}=\nu/\eta$ is 4 in this solution. 
The magnetic energy of the toroidal field  within the bulk of the convective zone 
is roughly an order of magnitude stronger than that of the poloidal field. 
Figure 6 displays a 3--D rendering of the toroidal field. Substantial magnetic helicity is 
present, involving complex winding of the toroidal magnetic fields along their length, 
with both polarities interchanging their position into structures not unlike cables. 
The toroidal magnetic fields also possess the greater spatial scales, having been stretched 
by the gradients in angular velocity. Some features could resemble magnetic flux tubes, 
although they are short lived. There is a clear north-south asymmetry in both the toroidal 
and poloidal magnetic field topology.

\begin{figure}[!ht]
\setlength{\unitlength}{1.0cm}
\begin{picture}(5,7.1)
\includegraphics{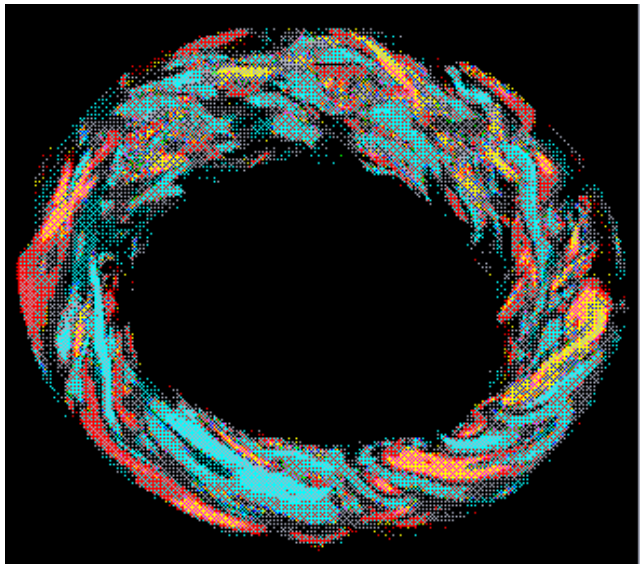}
\end{picture}
\caption[]{\label{fig6} Snapshot of the toroidal component of the magnetic field for case $M3$ 
appearing here as 3--D torus centered around the equator and slightly tilted forward.}
\end{figure}

\subsection{Maintaining a Strong Solar-like Differential Rotation}

With fairly strong magnetic fields sustained within the bulk of the convection 
zone in case $M3$, it is to be expected that the differential rotation $\Omega$ 
will respond to the feedback from the Lorentz forces. Figure 7 shows the time averaged 
angular velocity achieved in case $M3$. As for the convective motions, the main effect 
of the Lorentz forces is to extract energy from the kinetic energy stored in the 
differential rotation. The reduction of KE contained in the angular velocity is of the
same order as the decrease seen in the total KE, i.e. 40\%. As a consequence 
the angular velocity contrast $\Delta\Omega$ from 60$^\circ$ to the equator drops 
by $\sim 30\%$ in case $M3$, going from 140 nHz (or 34\% compared to the reference 
frame $\Omega_o$) in the hydrodynamic case to 100 nHz (or 24\%). Nevertheless,
the angular velocity in case $M3$ remains in good agreement with the solar profile,
both in amplitude and profile, as can been seen by comparing Fig. 1 and 7.
The source of the reduction of the latitudinal contrast of $\Omega$ can be attributed 
to the poleward transport of angular momentum by the Maxwell stresses (the mean magnetic fields having
a negligable contribution). Now the Reynolds stresses again need to balance the angular momentum transport 
by the meridional circulation, the viscous diffusion and the Maxwell stresses.
This leads to a less efficient speeding up of the equatorial regions. Since ME is only
7\% of KE in case $M3$, the Maxwell stresses are not yet the main players in
redistributing the angular momentum. We have found that a value of ME above about 20\% of
KE leads to a significant magnetic braking effect on the differential rotation. Had the
simulation been restarted with a stronger initial magnetic field $B_0$, $\Delta\Omega$ could
drop by 90\% in less than a few hundreds days, thus being at variance with helioseismic findings.
By letting the convective motions gradually adapt themself to the nonlinear feedback of 
the Lorentz forces one seems to limit the equipartitioning of ME and KE and thus succesfully
retain the strong differential rotation seen in pure hydrodynamical cases.

\begin{figure}[!ht]
\setlength{\unitlength}{1.0cm}
\begin{picture}(5,6.4)
\includegraphics{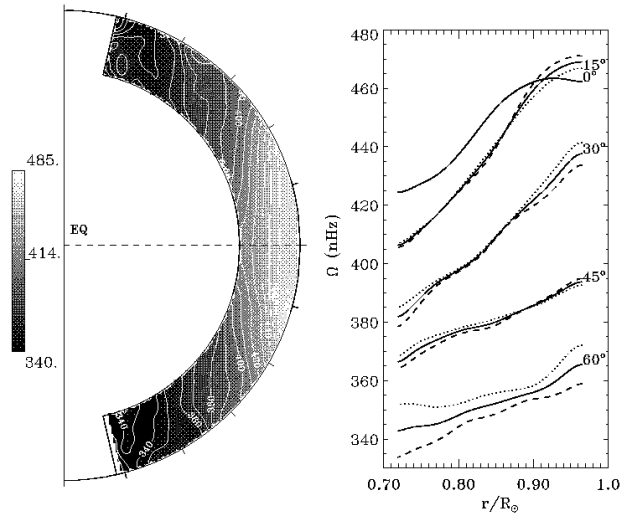}
\end{picture}
\caption[]{\label{fig7} Temporal and longitudinal averages of the angular velocity profiles
achieved in case $M3$ over an interval of 80 days. This case exhibits a prograde
equatorial rotation and a strong contrast $\Delta \Omega$ from equator to pole, as well as
possessing a high latitude region of particularly slow rotation. In the right panel a sense
of the asymmetry present in the solution can be assessed in these radial cuts at indicated latitudes.}
\end{figure}

Our 3--D simulations of convection in deep spherical shells, achieved through use of
massively parallel supercomputers, are helping to show how the strong differential
rotation present in the Sun may be maintained through fairly complex redistribution
of angular momentum by the turbulent compressible flows. We have also begun to study
the interaction of convection and rotation with seed magnetic fields in such shells,
thereby identifying some parameter ranges in which sustained magnetic dynamo action
can be realized without unduly reducing the angular velocity contrasts maintained by the
convection.

We thank Nicholas Brummell, Marc DeRosa, Peter Gilman,
Mark Miesch, Annick Pouquet and Jean-Paul Zahn for useful discussions.  
This work was partly supported by NASA through SEC Theory Program grant NAG5-8133
and by NSF through grant ATM-9731676.  Various phases of the
simulations with ASH were carried out with NSF PACI support of the San
Diego Supercomputer Center (SDSC), the National Center for
Supercomputing Applications (NCSA), and the Pittsburgh Supercomputing
Center (PSC).  Much of the analysis of the extensive data sets was
carried out in the Laboratory for Computational Dynamics (LCD) within
JILA.

\end{document}